\documentclass[a4paper]{article}

\usepackage{INTERSPEECH2022}
\usepackage{multirow}
\usepackage{subfigure}

\makeatletter
\renewcommand{\maketag@@@}[1]{\hbox{\m@th\normalsize\normalfont#1}}%
\makeatother

\title{Multichannel Speech Separation with Narrow-band Conformer 
\thanks{This work was supported by Zhejiang Provincial Natural Science Foundation of China under Grant 2022XHSJJ008. We thank Westlake University HPC Center for compution support.}
}
\name{Changsheng Quan$^{1,2}$, Xiaofei Li$^{2,*}$\thanks{* corresponding author}}
\address{
  $^1$Zhejiang University, Hangzhou, China\\
  $^2$Westlake University \& Westlake Institute for Advanced Study, Hangzhou, China}
\email{quanchangsheng@westlake.edu.cn,  lixiaofei@westlake.edu.cn}

\UseRawInputEncoding
\begin{document}

\maketitle
\begin{abstract}
This work proposes a multichannel speech separation method with narrow-band Conformer (named NBC). 
The network is trained to learn to automatically exploit narrow-band speech separation information, such as spatial vector clustering of multiple speakers. Specifically, in the short-time Fourier transform (STFT) domain, the network processes each frequency independently, and is shared by all frequencies. For one frequency, the network inputs the STFT coefficients of multichannel mixture signals, and predicts the STFT coefficients of separated speech signals.  Clustering of spatial vectors shares a similar principle with the self-attention mechanism in the sense of computing the similarity of vectors and then aggregating similar vectors. Therefore, Conformer would be especially suitable for the present problem. Experiments show that the proposed narrow-band Conformer achieves better speech separation performance than other state-of-the-art methods by a large margin. 

\end{abstract}

\section{Introduction}

Multichannel speech separation is to separate the speech signals for different speakers from the multichannel mixture signals.
The separated speech signals can then be used for better human hearing or other back-end tasks, such as automatic speech recognition and speaker identification.
Speech separation methods usually process the recorded signal in the short time Fourier transform (STFT) domain.
In the STFT domain, due to the sparsity of speech, the speech mixture satisfies the W-disjoint orthogonality assumption \cite{yilmaz_blind_2004}, namely we could approximately consider each time frequency (TF) point to be dominated by a single speaker.
The spatial features of one frequency, like steering vector and inter-channel phase difference (IPD), of frames dominated by the same speaker are consistent to each other, since they  originate from the same spatial location.
Leveraging the sparsity of speech and the consistency of spatial features, speech separation is proposed in \cite{winter_map-based_2006} by applying hierarchical clustering of the TF-wise mixing vectors.
Beamforming is a spatial filtering method, can also be used for speech separation.
The widely used minimum variance distortionless response (MVDR) beamformer preserves the speech with desired steering vector while suppresses other signal components \cite{gannot_consolidated_2017}.
In the CHiME-6 multispeaker speech recognition challenge \cite{watanabe_chime-6_2020}, Guided Source Separation (GSS) \cite{boeddecker_front-end_2018} method is widely adopted by participants, which combines the techniques of TF-bin clustering and MVDR beamforming, and achieves excellent speech separation performance.
Above methods \cite{winter_map-based_2006, gannot_consolidated_2017, boeddecker_front-end_2018} all perform frequency-wise speech separation in the STFT domain.

Deep learning has been first used for single-channel speech separation by learning the spectral pattern of speech \cite{hershey_deep_2016, yu_permutation_2017}.
When multiple microphones are available, in addition to the spectral pattern of speech, the spatial information of speakers can be employed, like inter-channel phase difference (IPD) \cite{wang_multi-channel_2018}, or automatically learned spatial features \cite{gu_enhancing_2020}.
Another way to leverage the spatial information is to directly predict the spatial filters \cite{luo_fasnet_2019, luo_end--end_2020} or to estimate the filter parameters using the separated speech signals of deep learning based methods, such as in \cite{ochiai_beam-tasnet_2020, heymann_blstm_2015}.
Recently, multi-head self-attention mechanism is proposed within the Transformer network \cite{vaswani_attention_2017}, to capture longer-term dependencies.
Later, Conformer \cite{gulati_conformer_2020} introduces convolutions to the Transformer architecture to leverage more local information, designed for automatic speech recognition.
\cite{chen_continuous_2021} adopts Conformer for continuous speech separation. Separation Transformer (Sepformer) \cite{subakan_attention_2021} modifies the Transformer architecture by adopting the dual-path mechanism proposed in \cite{li_dual-path_2021} to learn both short and long-term dependencies, and achieves outstanding signal-channel speech separation performance.

In this work, we propose a multichannel speech separation method with narrow-band Conformer, named NBC. This is a continuation work of our previous work \cite{quan_multi-channel_2021}, in which a bidirectional long short-term memory (BLSTM) network is proposed to perform multichannel speech separation in narrow-band, while this work adopts a more powerful Conformer network. In the STFT domain, the speech separation network processes each frequency independently, and is shared by all frequencies. For each frequency, the network takes as input the STFT coefficients of multichannel mixture signals, and predicts the STFT coefficients of separated speech sources.
It is known that one STFT frequency includes rich information for separating speech sources, such as the spectral sparsity of speech and the inter-channel differences of multiple sources. The proposed narrow-band network is trained to learn a function/rule to automatically exploit these information, and to perform end-to-end narrow-band speech separation. One important functional for narrow-band speech separation is to perform spatial vector clustering to group frames dominated by the same speaker, as is done in \cite{winter_map-based_2006}. Clustering of spatial vectors shares a similar principle with the self-attention mechanism in the sense of computing the similarity of vectors and then aggregating similar vectors. In addition, somehow narrow-band speech signal is a random process, thus local smoothing operations are often required for estimating speech statistics, e.g. cross correlation, which can be conducted by convolutional layers. Therefore, Conformer is expected to be especially suitable for multichannel narrow-band speech separation. Moreover, we reinforce the convolutional layers in Conformer to further improve the performance. 
Experimental results show that Conformer is indeed able to perform frame clustering, and the proposed model achieves better results than our previous work \cite{quan_multi-channel_2021}, FaSNet \cite{luo_fasnet_2019, luo_end--end_2020} and Sepformer \cite{subakan_attention_2021} by a large margin.
Code for the proposed method is available at \footnote{https://github.com/Audio-WestlakeU/NBSS}.

\section{Method}

This work is a continuation of our previous work of LSTM-based multichannel narrow-band speech separation \cite{quan_multi-channel_2021}. We replace the BLSTM network with a more powerful modified Conformer network. Multichannel narrow-band speech separation highly relies on the clustering of spatial (steering) vector of speakers, which shares a similar principle with the self-attention mechanism in the sense of computing the similarity of vectors and then aggregating similar vectors. In addition, narrow-band speech signal is a random process, thus local smoothing operations are often required for estimating speech statistics, e.g. cross correlation, which can be conducted by convolutional layers. Overall, Conformer would be especially suitable for multichannel narrow-band speech separation. 

\subsection{Recap of Narrow-band Speech Separation with fPIT}

We consider multichannel signals in the STFT domain: 
\begin{equation}
  {\rm X}_{f,t}^{m} = \sum_{n=1}^N {\rm Y}_{f,t}^{n,m},
  \label{eq1}
\end{equation}
where $f$ $\in$ $\{0,...,F-1\}$, $t$ $\in$ $\{1, ..., T\}$, $m$ $\in$ $\{1,...,M\}$ and $n\in \{1,...,N\}$ denote the indices of frequency, time frame, microphone channel and speaker, respectively.
${\rm X}_{f,t}^{m}$ and ${\rm Y}_{f,t}^{n,m}$ are the complex-valued STFT coefficients of microphone signals and of the reverberant spatial image of speech sources, respectively.
The target of this work is to recover the reverberant spatial images of multiple speakers at a reference channel, e.g.\ ${\rm Y}_{f,t}^{n,r}$, where $r$ denotes the index of the reference channel.

\begin{figure}[t]
  \centering
  \includegraphics[width=\linewidth]{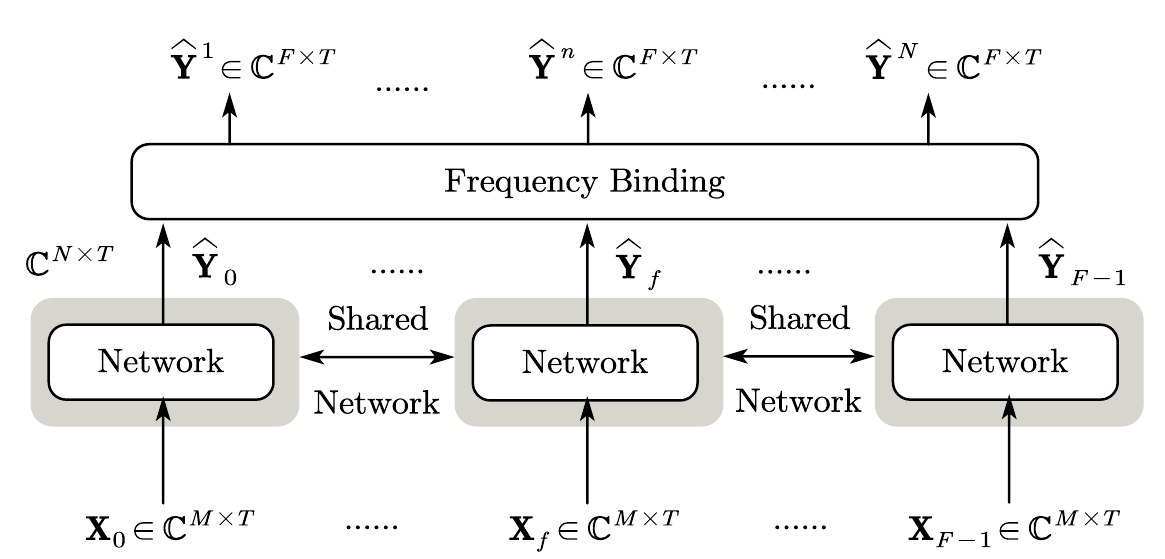}
  \caption{Framework of narrow-band deep speech separation.}
  \label{fig:flowchart}
\vspace{-0.6cm}
\end{figure}

The framework of our narrow-band method \cite{quan_multi-channel_2021} is shown in Fig.\ref{fig:flowchart}, where speech separation is performed independently for each frequency, and the same separation network is shared by all frequencies.
The network previously consists of two layers of BLSTM, and currently Conformer blocks, as will be presented in Section \ref{sec:conformer}. 
The STFT coefficients of a single frequency band $\mathbf{X}_{f}$ is taken as the network input sequence:
\begin{equation}
  \mathbf{X}_{f} = (\mathbf{X}_{f,1},\dots,\mathbf{X}_{f,T}) \in \mathbb{C}^{M\times T},
\end{equation}
where $\mathbf{X}_{f,t} = [{\rm X}_{f,t}^1,\dots,{\rm X}_{f,T}^1]^{T} \in \mathbb{C}^{M\times 1}$ concatenates the multichannel STFT coefficients of one TF bin. The network outputs the sequence of separated speech signals of the same frequency, denoted by $\boldsymbol{\widehat{{\rm Y}}}_{f}\in\mathbb{C}^{N\times T}$, which is the prediction of the ground truth signal $\boldsymbol{{{\rm Y}}}_{f}\in\mathbb{C}^{N\times T}$. For each time step, the output vector concatenates the STFT coefficients of $N$ speech sources. 
Note that the input and output sequence are denoted in the complex domain for presentation simplicity, while the real and imaginary parts of them are actually used in practice, with vector dimensions of $2M$ and $2N$, respectively.

To facilitate the network training, magnitude normalization is performed for each frequency, as $\mathbf{X}_{f}/ \overline{{\rm X}}_f$, where $\overline{{\rm X}}_{f} = \textstyle\sum_{t=1}^T |{\rm X}_{f,t}^r|/T$.
And, an inverse normalization is applied to the network output as $\boldsymbol{\widehat{{\rm Y}}}_{f} \overline{{\rm X}}_{f}$.

The frequency permutation problem is solved by a full-band permutation invariant training (fPIT) technique, which forces the network to produce predictions with the same speaker label permutation for all frequencies. 
Specifically, the predictions at the same output position of all frequencies, 
i.e.\ $\widehat{\mathbf{Y}}^n = [\widehat{\mathbf{Y}}_0^n;\dots;\widehat{\mathbf{Y}}_{F-1}^n]\in \mathbb{C}^{F\times T}$,
are forced to belong to the same speaker, and binded together to form the complete spectra of this speaker.
Thus, the best permutation for the $N$ bindings can be regarded as the permutation for all frequencies.
The loss of all frequencies can then be calculated in a fPIT way: 
\begin{equation}{
  \text{fPIT}(\boldsymbol{{\rm \widehat{Y}}}^{1},\ldots, \boldsymbol{{\rm \widehat{Y}}}^{N}, \boldsymbol{{\rm {Y}}}^{1},\ldots,\boldsymbol{{\rm {Y}}}^{N}) \\
  =
  \mathop{min}_{p\in \mathcal{P}}
  \sum_n \mathcal{L}(
    \boldsymbol{{\rm {Y}}}^n
    ,
    \boldsymbol{{\rm \widehat{Y}}}^{p(n)}
  ),
  \label{eq7}}
\end{equation} 
where $\boldsymbol{{\rm {Y}}}^n \in \mathbb{C}^{F\times T}$ is the ground truth speech of the $n$-th speaker, consisting of all the STFT coefficients (at the reference channel) of the $n$-th speaker. $\mathcal{P}$ denotes the set of all possible permutations, and
$p$ denotes a permutation in $\mathcal{P}$ which maps labels of ground truth to labels of predictions. $\mathcal{L}$ denotes a loss function.

SI-SDR \cite{roux_sdr_2019} is taken as the training loss function:
\begin{equation}
   \mathcal{L}(\mathbf{Y}^n,\mathbf{\widehat{\mathbf{Y}}}^{p(n)}) 
   = -10 \log_{10}\frac{\left \| \alpha \boldsymbol{{\rm y}}^n \right \|^2}{\left \| \alpha \boldsymbol{{\rm y}}^n - \boldsymbol{{\rm \widehat{y}}}^{p(n)} \right \|^2}
  \label{eq8}
\end{equation}
where $\alpha=({\boldsymbol{{\rm \widehat{y}}}^{p(n)}})^T\boldsymbol{{\rm y}}^n/\left \| \boldsymbol{{\rm y}}^n \right \|^2$, $\boldsymbol{{\rm y}}^n$
and $\boldsymbol{{\rm \widehat{y}}}^{p(n)}$ are the inverse STFT of $\boldsymbol{{\rm Y}}^n$ and $\boldsymbol{{\rm \widehat{Y}}}^{p(n)}$, respectively.

\subsection{Modified Conformer}
\label{sec:conformer}
\begin{figure}[t]
  \centering
  \includegraphics[width=\linewidth]{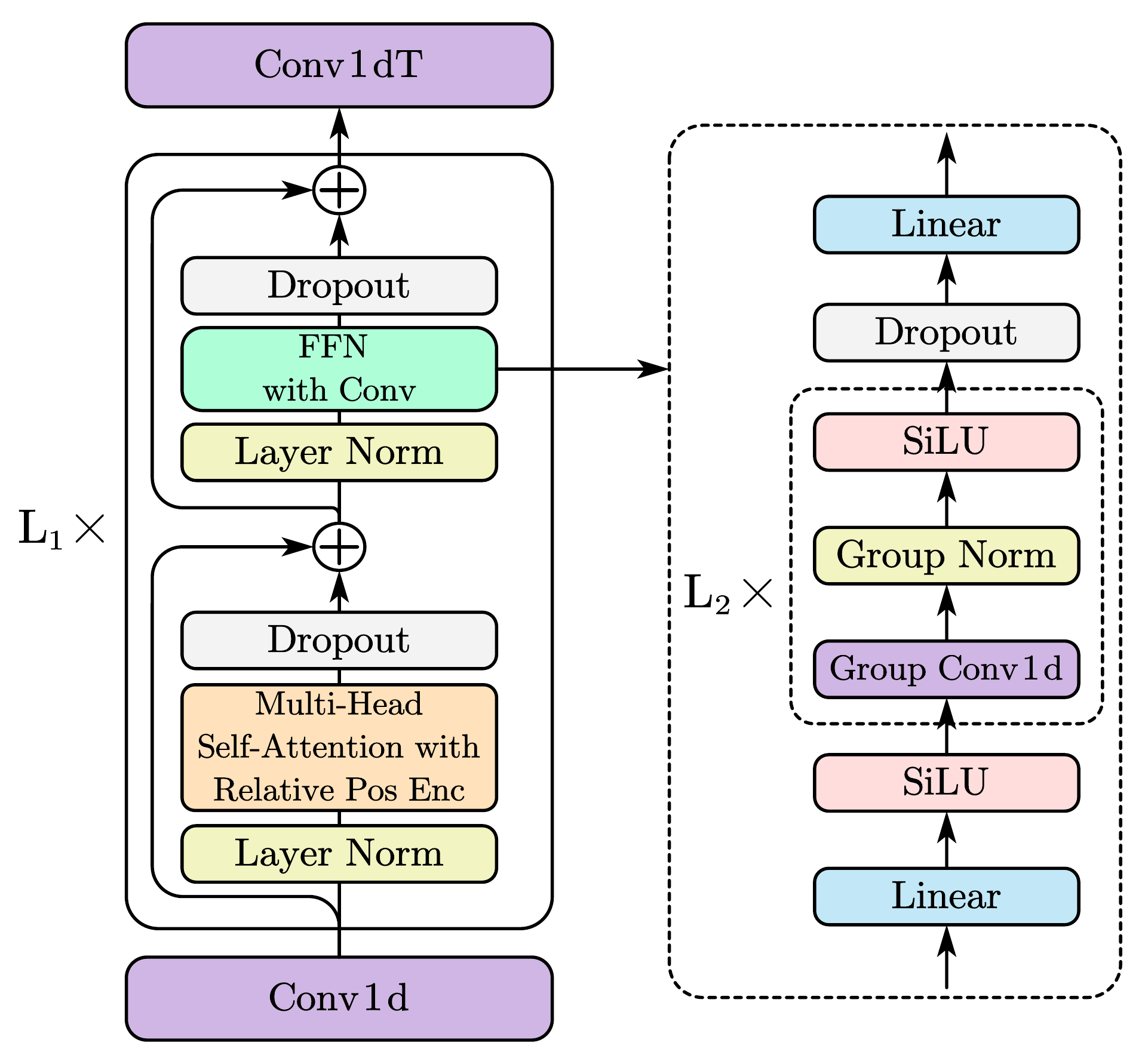}
  \caption{Architecture of the modified Conformer.}
  \label{fig:NBformer}
\vspace{-0.6cm}
\end{figure}

In this paper, we propose a modified Conformer to replace the BLSTM network used in our previous work. The network architecture is shown in Fig.~\ref{fig:NBformer}. It is composed of one convolution layer (Conv1d), $L_1$ modified Conformer blocks, and one transposed convolution layer (Conv1dT). Conv1d and Conv1dT are operated along the time dimension.
Conv1d takes as input $\boldsymbol{{\rm X}}_{f}$ and Conv1dT outputs $\widehat{\mathbf{Y}}_{f}$, thus the input channel of Conv1d and the output channel of Conv1dT are $2M$ and $2N$, respectively. The output of Conv1d layer, denoted as $x_0\in \mathbb{R}^{H_1 \times T}$, will be used as the input hidden representation of the following Conformer blocks, where $H_1$ is the number of hidden units.

The modified Conformer block is composed of two modules: multi-head self-attention with a relative positional encoding and feed forward network with convolutional layers, each of which is proceeded or followed by layer normalization, dropout or residual connection, as shown in Fig.~\ref{fig:NBformer}.

\subsubsection{Multi-head Self-attention with Relative Positional Encoding}

We employ the multi-head self-attention mechanism with a relative positional encoding technique introduced in \cite{dai_transformer-xl_2019}.
The relative positional encoding has been proven effective for automatic speech recognition  \cite{gulati_conformer_2020, pham_relative_2020}. Our preliminary experiments show that, for the present problem, relative positional encoding also outperforms absolute positional encoding \cite{vaswani_attention_2017}.

Denote the input of the $i$-th modified Conformer block as $x_{i-1}$, taking the residual connection into account, the output of the module of multi-head self-attention with relative positional encoding (referred to as RPSA) is:
\begin{equation}
  x^{'}_i = x_{i-1} + \text{ dropout(RPSA(LayerNorm($x_{i-1}$)))} \in \mathbb{R}^{H_1 \times T},
  \notag
\end{equation}
where LayerNorm means layer normalization.

\subsubsection{Feed Forward Network with Convolutions}

This module first increases the number of hidden units to $H_2$ using a linear layer:
\begin{equation}
  x^{''}_i =\text{ SiLU(Linear(LayerNorm($x^{'}_i$)))} \in \mathbb{R}^{H_2 \times T}, \\
  \notag
\end{equation}
where SiLU (Sigmoid Linear Unit) \cite{hendrycks_gaussian_2020}, also known as Swish \cite{ramachandran_searching_2017}, is the activation function.

Then, $L_2$ blocks of group convolution are employed to capture the local temporal information. Each block is composed of one 1D group convolution layer (GroupConv), one group normalization layer \cite{wu_group_2018} (GroupNorm) and one SiLU activation. The 1D group convolution is applied along the time dimension.
The $j$-th group convolution block of the $i$-th modified Conformer block can be formulated as:
\begin{equation}
  x^{'''}_{i,j} =\text{ SiLU(GroupNorm(GroupConv($x^{'''}_{i,j-1}$)))} \in \mathbb{R}^{H_2 \times T}, \\
  \notag
\end{equation}
where $j$ ranges from 1 to $L_2$, and $x^{'''}_{i,0} = x^{''}_i$.

At the end, another linear layer is used to reduce the number of hidden units back to $H_1$, and obtain the output of the $i$-th modified Conformer block:
\begin{equation}
  x_{i} =x^{'}_i + \text{ dropout(Linear(dropout($x^{'''}_{i,L_2}$)))} \in \mathbb{R}^{H_1 \times T}. \\
  \notag
\end{equation}
The output of the last modified Conformer block is converted to $\boldsymbol{\widehat{{\rm Y}}}_{f}$ by its following Conv1dT layer.

Overall, the major difference between the Conformer network proposed in  \cite{gulati_conformer_2020} and the modified Conformer used in this work is that we largely reinforce the convolution layers to account for the requirement of narrowband speech separation. Briefly speaking, we use more convolutional layers and more complex group convolution, and put the convolutional layers in between the two feed forward layers with $H_2$ channels. 

\section{Experiments}

\subsection{Experimental Setup}
\subsubsection{Dataset}
The proposed method is evaluated on a spatialized version of the WSJ0-2mix dataset \cite{hershey_deep_2016}, which contains 20000, 5000 and 3000 speech pairs for training, validation and test respectively.
In our experiments, each speech pair is overlapped in a manner where the tail part of one signal is overlapped with the head part of the other signal, which is also used in \cite{luo_end--end_2020, quan_multi-channel_2021}.
The overlap ratio is uniformly sampled from the range of [10\%, 100\%].
The resulting mixed utterances are all set to four-second long.
Room impulse responses are simulated using a GPU based implementation of the image method \cite{allen_image_1979}, called gpuRIR \cite{diaz-guerra_gpurir_2021}.
The length, width and height of the simulated rooms are uniformly sampled in the range of [3 m, 8 m], [3 m, 8 m] and [3 m, 4 m], respectively.
The reverberation time (RT60) of each room is uniformly sampled from the range of [0.1 s, 1.0 s].
A horizontal 8-channel circular microphone array with a radius of 5 cm is used.
The center of microphone array is randomly put in a square area (diameter is 1 m) located at the room center with a height of 1.5 m. 
Speaker locations are randomly sampled in the room with a height of 1.5 m and with the direction difference between two speakers randomly sampled from 0$^\circ$ to 180$^\circ$.
Each speaker is located at least 0.5 m away from the walls.

\subsubsection{Training Configurations}
According to our preliminary experiments, we set the kernel size of the first convolution layer and the last transposed convolution layer to 4, and the kernel size of group convolutions to 3.
The number of self-attention heads, and the number of groups for group convolution and group normalization are all set to 8.
The numbers of hidden units $H_1$ and $H_2$ are set to 192 and 384, respectively.
The numbers of modified Conformer blocks and group convolution blocks are set to $L_1=4$ and $L_2=3$, respectively.

The sampling rate is 16 kHz. STFT is applied to speech signals using a hanning window of length 512 samples (32ms) with a hop size of 256 samples.
The network is trained with 16 utterances per mini-batch, thus the batch size of narrow-band training samples (frequencies) is 4112 ($16 \times 257$).
The Adam \cite{kingma2015adam} optimizer is used with an initial learning rate of 0.001.
When the validation loss does not decrease in 3 consecutive epochs, the learning rate is halved until it reaches a given minimum of 0.0001.
Gradient clipping is applied with a threshold of 5.

\subsubsection{Performance Metrics and Comparison Methods}
To evaluate the speech separation performance, four metrics are used: i) narrow-band PESQ (NB-PESQ) \cite{rix_perceptual_2001}, ii) wide-band PESQ (WB-PESQ) \cite{rix_perceptual_2001}, and iii) SI-SDR \cite{roux_sdr_2019}, and iv) real time factor (RTF), which is tested on a personal computer equipped with Intel(R) i7-9700 CPU (3.0 GHz).

We compare the proposed method with four baselines. 
(a) Oracle MVDR beamformer \footnote{https://github.com/Enny1991/beamformers}, for which the steering vector of desired speech and the covariance matrix of undesired signals are computed using the true desired speech and undesired signals, respectively.
(b) FaSNet with TAC mechanism \cite{luo_end--end_2020}, referred to as FaSNet-TAC.
(c) SepFormer proposed in \cite{subakan_attention_2021}. Sepformer is a transformer-based single channel speech separation model. We modify Sepformer to account for the multichannel case, by simply changing the input channel of its first convolution layer from 1 to the number of microphone channels, referred to as SepFormerMC. Note that this change may not be the optimal way, but it still improves the performance of single-channel SepFormer.
(d) Our previously proposed narrow-band BLSTM method \cite{quan_multi-channel_2021}, referred to a NB-BLSTM.

\subsection{Experimental Results}

\subsubsection{Comparison}

\setlength\tabcolsep{3pt}
\begin{table}[t]
 \footnotesize
  \centering
  \caption{Speech separation results.}
  \label{tab1}
  \begin{tabular}{@{}cccccc@{}}
  \toprule
   & \#param & NB-PESQ & WB-PESQ  & SISDR & RTF \\
     & (M) & & & (dB) &  \\
  \midrule
  mixture
  & -     & 2.05 & 1.59 &  0.0 & -     \\
  Oracle MVDR 
  & -     & 3.16 & 2.65 &  11.0 & -     \\
  FaSNet-TAC  \cite{luo_end--end_2020} 
  & 2.8 & 2.96 & 2.53  & 12.6 &  0.67  \\
  SepFormer   \cite{subakan_attention_2021} 
  & 25.7 & 3.17 & 2.72  & 13.2 & 1.69   \\
  SepFormerMC 
  & 25.7 & 3.42 & 3.01  & 14.9 & 1.70   \\
  \hline
  NB-BLSTM \cite{quan_multi-channel_2021}
  & 1.2 & 3.28 & 2.81  & 12.8 &  0.37    \\
 prop.
  & 2.0 & \textbf{4.00}  & \textbf{3.78}  & \textbf{20.3}  &   1.32   \\
  \bottomrule
  \end{tabular}
\end{table}

\begin{table}[t]
\centering
  \caption{Speech separation results with various number of convolutional layers in the present narrow-band framework.} 
  \label{tab:ablation}
  \footnotesize
  \begin{tabular}{cccccc}
  \toprule
            & \#param & NB-PESQ   & WB-PESQ     & SISDR   \\
  \# Convolutional layers     & (M) & & & (dB)    \\
  \midrule     
0     & 1.4    & 3.66      & 3.29      & 16.4   \\
2     & 1.8    & 3.92      & 3.66      & 19.2   \\
3     & 2.0    &\textbf{4.00}&\textbf{3.78}&\textbf{20.3}  \\
4     & 2.3    & 3.99      & 3.76      & 20.1   \\
Conformer \cite{gulati_conformer_2020}     & 2.4    & 3.94      & 3.69     & 19.1   \\%
  \bottomrule
  \end{tabular}
\vspace{-0.6cm}
\end{table}

Table \ref{tab1} shows the comparison results, where the number of parameters is also reported. All deep-learning-based methods obtain better SI-SDR scores than MVDR, which indicates that deep neural network is able to better separate the mixture signal. The single channel SepFormer \cite{subakan_attention_2021} achieves outstanding performance, which testifies the capability of transformer for speech separtion. By using more microphone channels, SepFormerMC further noticeably improves the performance. Our previously proposed narrowband BLSTM already achieves competitive performance, which proves the effectiveness of the proposed narrow-band framework that training a network to perform narrow-band speech separation by learning to discriminate the spatial cues of different speakers. By replacing BLSTM with our modified Conformer, the performance is further significantly improved, which verifies our assertion that Conformer is especially fit for narrow-band speech separation, in terms of spatial vector clustering and signal smoothing. Overall, the proposed method outperforms other methods by a very large margin. Moreover, this excellent performance is achieved with a very small network, as the numbers of hidden units are $H_1$=192 and $H_2$=384, and total number of parameters are 2.0 M. Since the proposed narrow-band network needs to run for all frequencies independently, its RTF (=1.32) cannot reach the real time requirement yet, which can be easily solved by further reduce the network size as for future works. Some audio examples can be found in our webpage \footnote{https://audio.westlake.edu.cn/Research/nbss.htm}.

\subsubsection{Ablation Study}
We conduct ablation studies to evaluate the effectiveness of convolution layers employed in each block of the modified Conformer. The results are reported in Table~\ref{tab:ablation}. The results of using the Conformer architecture presented in \cite{gulati_conformer_2020} are also given, note that this Conformer architecture is evaluated in the present narrow-band framework and uses the same parameter settings as the proposed network in term of the number of blocks and the number of hidden units for each block.  
It can be seen that the performance measures rapidly increase along with the increasing of convolutional layers up to 3 layers, which indicates that convolution plays an important role for narrow-band speech separation. Compared to Conformer, the modified Conformer of this work achieves better performance with less parameters. 

\subsubsection{Attention Analysis}

\begin{figure}[t]
  \centering
  \includegraphics[width=0.95\linewidth]{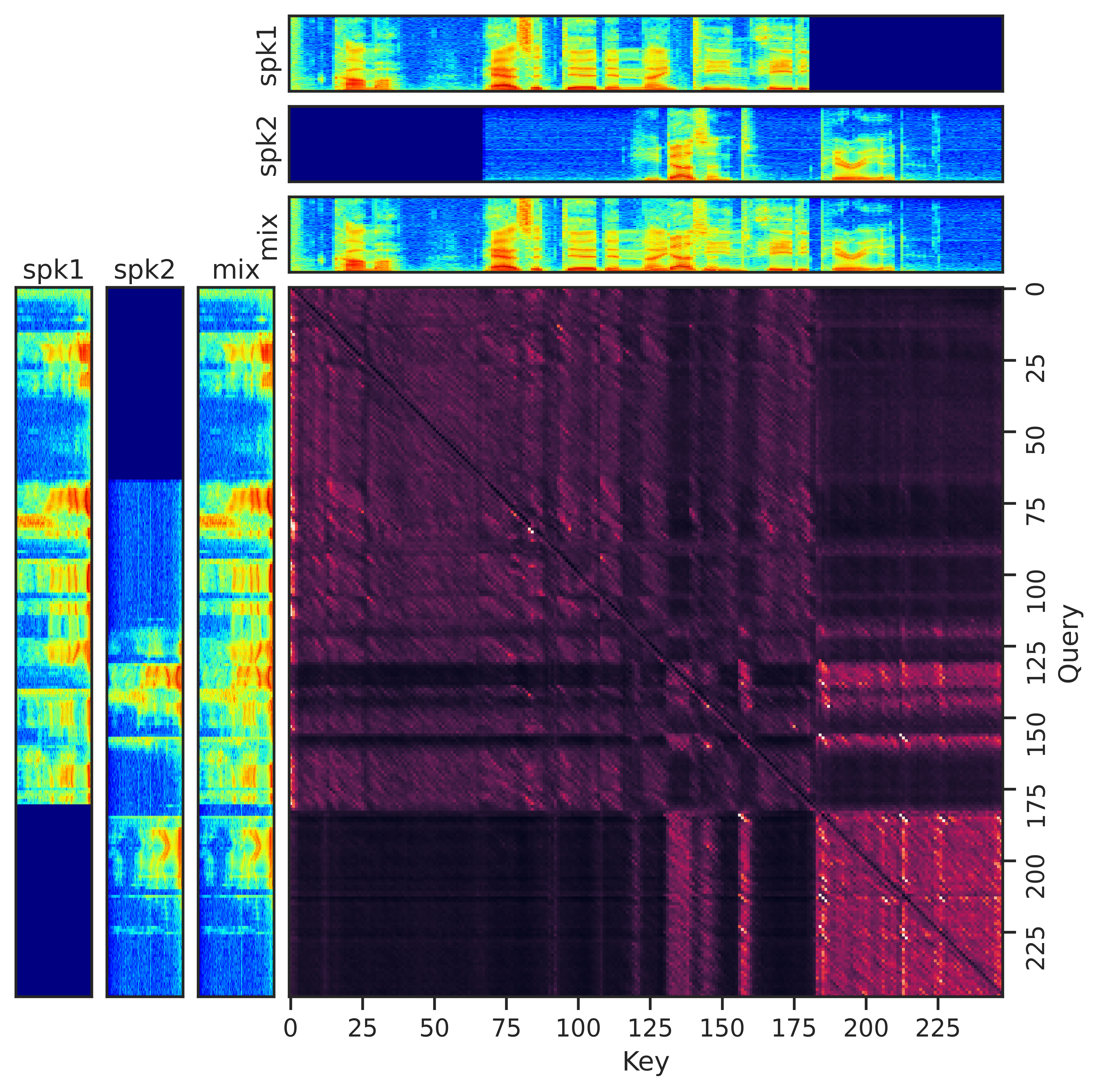}
  \caption{An attention map of one utterance. 'spk1' and 'spk2' stands for the clean speech of the first and second speaker, respectively. Attentions are averaged over all frequencies.}
  \label{fig:attn}
  \vspace{-0.2cm}
\end{figure}

Fig.~\ref{fig:attn} shows an attention map obtained by the $8$-th head of the $2$-th block, for one example utterance, where attentions are averaged over all frequencies. It can be seen that the frames dominated by the same speaker attend to each other, thus this attention map is clustered into different groups by speakers. This exactly meets our expectations that the self-attention mechanism can perform frame clustering, possibly based on the spatial vector of frames.

\section{Conclusions}

This work has proposed a multichannel speech separation method with narrow-band Conformer, named NBC. 
The network learns to automatically exploit narrow-band information to perform speech separation, such as based on spatial vector clustering, while the self-attention mechanism is especially suitable for this task. Experiments show that the proposed narrow-band Conformer is indeed able to clustering the frames associated to the same speaker. Overall, the proposed method achieves the state-of-the-art performance, and surpasses other methods by a large margin.   

\newpage

\bibliographystyle{IEEEtran}

\bibliography{mybib}

\begin{thebibliography}{10}
\providecommand{\url}[1]{#1}
\csname url@samestyle\endcsname
\providecommand{\newblock}{\relax}
\providecommand{\bibinfo}[2]{#2}
\providecommand{\BIBentrySTDinterwordspacing}{\spaceskip=0pt\relax}
\providecommand{\BIBentryALTinterwordstretchfactor}{4}
\providecommand{\BIBentryALTinterwordspacing}{\spaceskip=\fontdimen2\font plus
\BIBentryALTinterwordstretchfactor\fontdimen3\font minus
  \fontdimen4\font\relax}
\providecommand{\BIBforeignlanguage}[2]{{%
\expandafter\ifx\csname l@#1\endcsname\relax
\typeout{** WARNING: IEEEtran.bst: No hyphenation pattern has been}%
\typeout{** loaded for the language `#1'. Using the pattern for}%
\typeout{** the default language instead.}%
\else
\language=\csname l@#1\endcsname
\fi
#2}}
\providecommand{\BIBdecl}{\relax}
\BIBdecl

\bibitem{yilmaz_blind_2004}
O.~Yilmaz and S.~Rickard, ``Blind {Separation} of {Speech} {Mixtures} via
  {Time}-{Frequency} {Masking},'' \emph{IEEE Transactions on Signal
  Processing}, vol.~52, no.~7, pp. 1830--1847, Jul. 2004.

\bibitem{winter_map-based_2006}
S.~Winter, W.~Kellermann, H.~Sawada, and S.~Makino, ``{MAP}-{Based}
  {Underdetermined} {Blind} {Source} {Separation} of {Convolutive} {Mixtures}
  by {Hierarchical} {Clustering} and $l_1$-{Norm} {Minimization},''
  \emph{EURASIP Journal on Advances in Signal Processing}, vol. 2007, no.~1,
  pp. 1--12, Dec. 2006.

\bibitem{gannot_consolidated_2017}
S.~Gannot, E.~Vincent, S.~Markovich-Golan, and A.~Ozerov, ``A {Consolidated}
  {Perspective} on {Multimicrophone} {Speech} {Enhancement} and {Source}
  {Separation},'' \emph{IEEE/ACM Transactions on Audio, Speech, and Language
  Processing}, vol.~25, no.~4, pp. 692--730, Apr. 2017.

\bibitem{watanabe_chime-6_2020}
S.~Watanabe, M.~Mandel, J.~Barker, E.~Vincent, A.~Arora, X.~Chang,
  S.~Khudanpur, V.~Manohar, D.~Povey, D.~Raj, and {others}, ``{CHiME}-6
  challenge: {Tackling} multispeaker speech recognition for unsegmented
  recordings,'' \emph{arXiv preprint arXiv:2004.09249}, 2020.

\bibitem{boeddecker_front-end_2018}
C.~Boeddecker, J.~Heitkaemper, J.~Schmalenstroeer, L.~Drude, J.~Heymann, and
  R.~Haeb-Umbach, ``Front-end processing for the {CHiME}-5 dinner party
  scenario,'' in \emph{CHiME}, Sep. 2018, pp. 35--40.

\bibitem{hershey_deep_2016}
J.~R. Hershey, Z.~Chen, J.~Le~Roux, and S.~Watanabe, ``Deep clustering:
  {Discriminative} embeddings for segmentation and separation,'' in
  \emph{ICASSP}, Mar. 2016, pp. 31--35.

\bibitem{yu_permutation_2017}
D.~Yu, M.~Kolbaek, Z.-H. Tan, and J.~Jensen, ``Permutation invariant training
  of deep models for speaker-independent multi-talker speech separation,'' in
  \emph{ICASSP}, Mar. 2017, pp. 241--245.

\bibitem{wang_multi-channel_2018}
Z.-Q. Wang, J.~Le~Roux, and J.~R. Hershey, ``Multi-{Channel} {Deep}
  {Clustering}: {Discriminative} {Spectral} and {Spatial} {Embeddings} for
  {Speaker}-{Independent} {Speech} {Separation},'' in \emph{ICASSP}, Apr. 2018,
  pp. 1--5.

\bibitem{gu_enhancing_2020}
R.~Gu, S.-X. Zhang, L.~Chen, Y.~Xu, M.~Yu, D.~Su, Y.~Zou, and D.~Yu,
  ``Enhancing {End}-to-{End} {Multi}-{Channel} {Speech} {Separation} {Via}
  {Spatial} {Feature} {Learning},'' in \emph{ICASSP}, May 2020, pp. 7319--7323.

\bibitem{luo_fasnet_2019}
Y.~Luo, C.~Han, N.~Mesgarani, E.~Ceolini, and S.-C. Liu, ``{FaSNet}:
  {Low}-{Latency} {Adaptive} {Beamforming} for {Multi}-{Microphone} {Audio}
  {Processing},'' in \emph{ASRU}, 2019, pp. 260--267.

\bibitem{luo_end--end_2020}
Y.~Luo, Z.~Chen, N.~Mesgarani, and T.~Yoshioka, ``End-to-end {Microphone}
  {Permutation} and {Number} {Invariant} {Multi}-channel {Speech}
  {Separation},'' in \emph{ICASSP}, May 2020, pp. 6394--6398.

\bibitem{ochiai_beam-tasnet_2020}
T.~Ochiai, M.~Delcroix, R.~Ikeshita, K.~Kinoshita, T.~Nakatani, and S.~Araki,
  ``Beam-{TasNet}: {Time}-domain {Audio} {Separation} {Network} {Meets}
  {Frequency}-domain {Beamformer},'' in \emph{ICASSP}, May 2020, pp.
  6384--6388.

\bibitem{heymann_blstm_2015}
J.~Heymann, L.~Drude, A.~Chinaev, and R.~Haeb-Umbach, ``{BLSTM} supported {GEV}
  beamformer front-end for the {3RD} {CHiME} challenge,'' in \emph{ASRU}, 2015,
  pp. 444--451.

\bibitem{vaswani_attention_2017}
A.~Vaswani, N.~Shazeer, N.~Parmar, J.~Uszkoreit, L.~Jones, A.~N. Gomez,
  Å.~Kaiser, and I.~Polosukhin, ``Attention is {All} you {Need},'' in
  \emph{Advances in {Neural} {Information} {Processing} {Systems}}, vol.~30,
  2017, pp. 5998--6008.

\bibitem{gulati_conformer_2020}
A.~Gulati, J.~Qin, C.-C. Chiu, N.~Parmar, Y.~Zhang, J.~Yu, W.~Han, S.~Wang,
  Z.~Zhang, Y.~Wu, and R.~Pang, ``Conformer: {Convolution}-augmented
  {Transformer} for {Speech} {Recognition},'' in \emph{Interspeech}, Oct. 2020,
  pp. 5036--5040.

\bibitem{chen_continuous_2021}
S.~Chen, Y.~Wu, Z.~Chen, J.~Wu, J.~Li, T.~Yoshioka, C.~Wang, S.~Liu, and
  M.~Zhou, ``Continuous {Speech} {Separation} with {Conformer},'' in
  \emph{ICASSP}, Jun. 2021, pp. 5749--5753.

\bibitem{subakan_attention_2021}
C.~Subakan, M.~Ravanelli, S.~Cornell, M.~Bronzi, and J.~Zhong, ``Attention {Is}
  {All} {You} {Need} {In} {Speech} {Separation},'' in \emph{ICASSP}, 2021, pp.
  21--25.

\bibitem{li_dual-path_2021}
C.~Li, Y.~Luo, C.~Han, J.~Li, T.~Yoshioka, T.~Zhou, M.~Delcroix, K.~Kinoshita,
  C.~Boeddeker, Y.~Qian, S.~Watanabe, and Z.~Chen, ``Dual-{Path} {RNN} for
  {Long} {Recording} {Speech} {Separation},'' in \emph{2021 {IEEE} {Spoken}
  {Language} {Technology} {Workshop} ({SLT})}, Jan. 2021, pp. 865--872.

\bibitem{quan_multi-channel_2021}
C.~Quan and X.~Li, ``Multi-channel {Narrow-band} {Deep} {Speech} {Separation}
  with {Full-band} {Permutation} {Invariant} {Training},'' in \emph{ICASSP},
  May 2022.

\bibitem{roux_sdr_2019}
J.~L. Roux, S.~Wisdom, H.~Erdogan, and J.~R. Hershey, ``{SDR} – {Half}-baked
  or {Well} {Done}?'' in \emph{ICASSP}, May 2019, pp. 626--630.

\bibitem{dai_transformer-xl_2019}
Z.~Dai, Z.~Yang, Y.~Yang, J.~Carbonell, Q.~Le, and R.~Salakhutdinov,
  ``Transformer-{XL}: {Attentive} {Language} {Models} beyond a {Fixed}-{Length}
  {Context},'' in \emph{Proceedings of the 57th {Annual} {Meeting} of the
  {Association} for {Computational} {Linguistics}}, 2019, pp. 2978--2988.

\bibitem{pham_relative_2020}
N.-Q. Pham, T.-L. Ha, T.-N. Nguyen, T.-S. Nguyen, E.~Salesky, S.~Stüker,
  J.~Niehues, and A.~Waibel, ``Relative {Positional} {Encoding} for {Speech}
  {Recognition} and {Direct} {Translation},'' in \emph{Interspeech}, Oct. 2020,
  pp. 31--35.

\bibitem{hendrycks_gaussian_2020}
D.~Hendrycks and K.~Gimpel, ``Gaussian {Error} {Linear} {Units} ({GELUs}),''
  \emph{arXiv preprint arXiv:1606.08415}, Jul. 2020.

\bibitem{ramachandran_searching_2017}
P.~Ramachandran, B.~Zoph, and Q.~V. Le, ``Searching for {Activation}
  {Functions},'' \emph{arXiv preprint arXiv:1710.05941}, Oct. 2017.

\bibitem{wu_group_2018}
Y.~Wu and K.~He, ``Group {Normalization},'' \emph{arXiv preprint
  arXiv:1803.08494}, Jun. 2018.

\bibitem{allen_image_1979}
J.~B. Allen and D.~A. Berkley, ``Image method for efficiently simulating
  small‐room acoustics,'' \emph{The Journal of the Acoustical Society of
  America}, vol.~65, no.~4, pp. 943--950, Apr. 1979.

\bibitem{diaz-guerra_gpurir_2021}
D.~Diaz-Guerra, A.~Miguel, and J.~R. Beltran, ``{gpuRIR}: {A} python library
  for room impulse response simulation with {GPU} acceleration,''
  \emph{Multimedia Tools and Applications}, vol.~80, no.~4, pp. 5653--5671,
  Feb. 2021.

\bibitem{kingma2015adam}
D.~P. {Kingma} and J.~L. {Ba}, ``Adam: A method for stochastic optimization,''
  in \emph{International Conference on Learning Representations}, 2015.

\bibitem{rix_perceptual_2001}
A.~Rix, J.~Beerends, M.~Hollier, and A.~Hekstra, ``Perceptual evaluation of
  speech quality ({PESQ})-a new method for speech quality assessment of
  telephone networks and codecs,'' in \emph{ICASSP}, 2001, pp. 749--752.

\end{thebibliography}


\end{document}